\documentstyle[12pt,preprint]{aastex}  

\begin{document}

\title{Beyond the Kuiper Belt Edge: New High Perihelion Trans-Neptunian Objects With Moderate Semi-major Axes and Eccentricities}
\author{Scott S. Sheppard\altaffilmark{1}, Chadwick Trujillo\altaffilmark{2} and David J. Tholen\altaffilmark{3}}

\altaffiltext{1}{Department of Terrestrial Magnetism, Carnegie Institution for Science, 5241 Broad Branch Rd. NW, Washington, DC 20015, USA, ssheppard@carnegiescience.edu}
\altaffiltext{2}{Gemini Observatory, 670 North A`ohoku Place, Hilo, HI 96720, USA}
\altaffiltext{3}{Institute for Astronomy, University of Hawai'i, Honolulu, HI 96822, USA}

\begin{abstract}  

We are conducting a survey for distant solar system objects beyond the
Kuiper Belt edge ($\sim 50$ AU) with new wide-field cameras on the
Subaru and CTIO telescopes.  We are interested in the orbits of
objects that are decoupled from the giant planet region in order to
understand the structure of the outer solar system, including whether
a massive planet exists beyond a few hundred AU as first reported in
Trujillo and Sheppard (2014).  In addition to discovering extreme
trans-Neptunian objects detailed elsewhere, we have found several
objects with high perihelia ($q>40$ AU) that differ from the extreme
and inner Oort cloud objects due to their moderate semi-major axes
($50<a<100$ AU) and eccentricities ($e\lesssim 0.3$).  Newly
discovered objects 2014 FZ71 and 2015 FJ345 have the third and fourth
highest perihelia known after Sedna and 2012 VP113, yet their orbits
are not nearly as eccentric or distant.  We found several of these
high perihelion but moderate orbit objects and observe that they are
mostly near Neptune mean motion resonances and have significant
inclinations ($i>20$ degrees).  These moderate objects likely obtained
their unusual orbits through combined interactions with Neptune's mean
motion resonances and the Kozai resonance, similar to the origin
scenarios for 2004 XR190.  We also find the distant 2008 ST291 has
likely been modified by the MMR+KR mechanism through the 6:1 Neptune
resonance.  We discuss these moderately eccentric, distant objects
along with some other interesting low inclination outer classical belt
objects like 2012 FH84 discovered in our ongoing survey.

\end{abstract}


\section{Introduction}

The Kuiper Belt is composed of small icy bodies just beyond Neptune.
It has been dynamically and collisionally processed (Morbidelli et
al. 2008; Petit et al. 2011).  Much of the structure of the Kuiper
Belt can be explained through interactions with Neptune (Dawson and
Murray-Clay 2012; Nesvorny 2015a,2015b; Nesvorny and Vokrouhlicky
2016).  The Neptune resonant objects were likely emplaced by Neptune's
outward migration (Malhotra 1995; Gomes et al. 2005; Gladman et
al. 2012; Sheppard 2012).  The scattered objects not in resonance have
large eccentricities with perihelia near Neptune ($q<38$ AU)
suggesting strong interactions with the planet (Gomes et al. 2008;
Brasil et al. 2014a).  Extreme trans-Neptunian (ETNOs) or inner Oort
cloud objects have high perihelia ($q>40$ AU), large semi-major axes
($a>150$ AU) and large eccentricities (Gladman et al. 2002; Morbidelli
and Levison 2004; Brown et al. 2004; Gomes et al. 2005,2006; Trujillo
and Sheppard 2014).  These extreme objects are currently decoupled
from the giant planets but must have interacted with something in the
past to obtain their extreme orbits (Kenyon and Bromley 2004; Gladman
and Chan 2006; Schwamb et al. 2010; Brasser et al. 2012; Soares and
Gomes 2013).  The similarity in the extreme objects' orbital angles
suggests they are being shepherded by an unseen massive distant planet
(Trujillo and Sheppard 2014; Batygin and Brown 2016).  There is an
edge to the Kuiper Belt for low to moderately eccentric objects around
48 AU (Jewitt et al. 1998; Trujillo and Brown 2001; Allen et
al. 2002).

Early dynamical simulations showed objects scattered by Neptune could
obtain high perihelia, moderately eccentric orbits from Neptune
interactions (Torbett \& Smoluchowski 1990; Holman and Wisdom 1993;
Malhotra 1995).  Until now, only one object, 2004 XR190, was known to
have a perihelion significantly beyond the Kuiper Belt edge yet only
have a moderate eccentricity and moderate semi-major axis (Allen et
al. 2006).  2004 XR190 likely obtained its high perihelion during
Neptune's outward migration, where the combined effect of the 8:3
Neptune Mean Motion Resonance (MMR) along with the Kozai (or
Lidov-Kozai) Resonance (KR) modified the eccentricity and inclination
of 2004 XR190 to obtain a very high perihelion (Gomes et al. 2008;
Gomes 2011).  The MMR+KR high perihelion objects may allow insights
into the past migrational history of Neptune.  In this letter we
report several new high perihelia objects ($q>40$ AU) that have only
moderate eccentricities ($e\lesssim 0.3$) and semi-major axes
($50<a<100$ AU) showing a significant population of these objects
exist.  This work is part of our ongoing survey and here we focus on
the moderate objects found beyond the Kuiper Belt edge.

\section{Observations}

Basic methodology of the survey have been published in Trujillo and
Sheppard (2014) and further details will be published elsewhere
(Sheppard and Trujillo in prep).  The majority of the area surveyed
was with the CTIO 4m Blanco telescope in Chile with the 2.7 square
degree Dark Energy Camera (DECam).  DECam has 62 $2048\times 4096$
pixel CCD chips with a scale of 0.26 arcseconds per pixel (Flaugher et
al. 2015).  The r-band filter was used during the early observing runs
(November and December 2012 and March, May and November 2013) reaching
to about 24th magnitude while the wide VR filter was used in the later
observations (March and September 2014 and April 2015) to about 24.5
magnitudes.  In addition, we have used the Subaru 8m telescope in
Hawaii with its 1.5 square degree HyperSuprimeCam.  HyperSuprimeCam
has 110 CCD chips with scale of 0.17 arcseconds per pixel.  The
observations were obtained in March and May 2015 to just over 25th
magnitude in the r-band.  We covered 1078 and 72 square degrees at
CTIO and Subaru, respectively, for a total of 1150 square degrees.

Most fields had three images of similar depth obtained over 3 to 6
hours.  Observations were within 1.5 hours of opposition, which means
the dominant apparent motion would be parallactic, and thus inversely
related to distance.  The seeing was between 0.6 and 1.2 arcseconds
for most fields allowing us to detect objects moving faster than 0.28
arcseconds per hour, which corresponds to about 500 AU at opposition,
though many fields would have detected objects to over 1000 AU
(determined by placing artificial slow objects in the fields).
Anything discovered beyond 50 AU was flagged for future recovery.
Most of the survey fields were between 5 and 20 degrees from the
ecliptic with fairly uniform longitudinal coverage.

\section{Results}

\begin{figure}
\epsscale{0.4}
\centerline{\includegraphics[angle=90,totalheight=0.6\textheight]{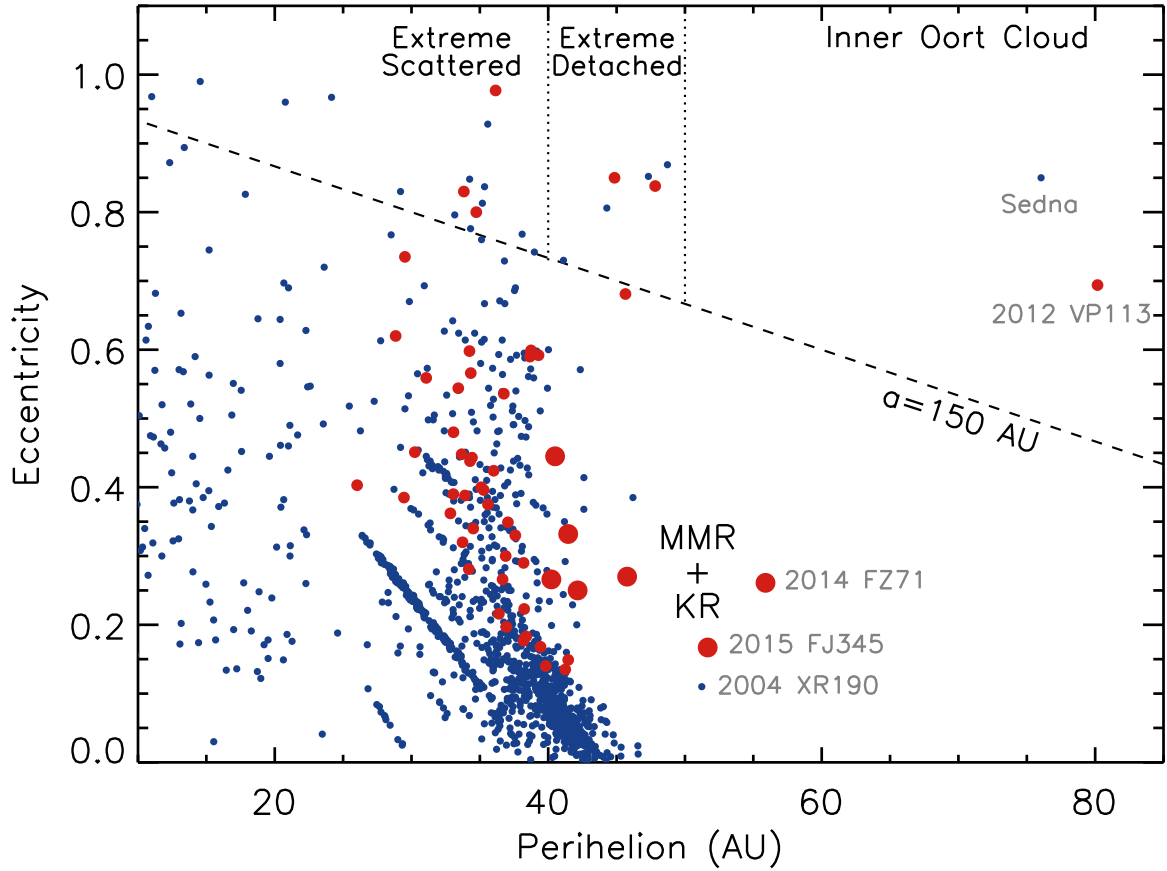}}
\caption{The perihelion versus eccentricity.  Red circles are objects
  discovered during this survey (large red circles are the focus of
  this work).  Objects above the dashed line are considered extreme
  with $a>150$ AU.  Objects with high perihelia beyond the Kuiper
  Belt edge at 50 AU but only moderate eccentricity are likely created
  by a combination of Neptune Mean Motion Resonances (MMR) and the
  Kozai Resonance (KR).}
\label{fig:kboeq2016} 
\end{figure}

\begin{figure}
\epsscale{0.4}
\centerline{\includegraphics[angle=90,totalheight=0.6\textheight]{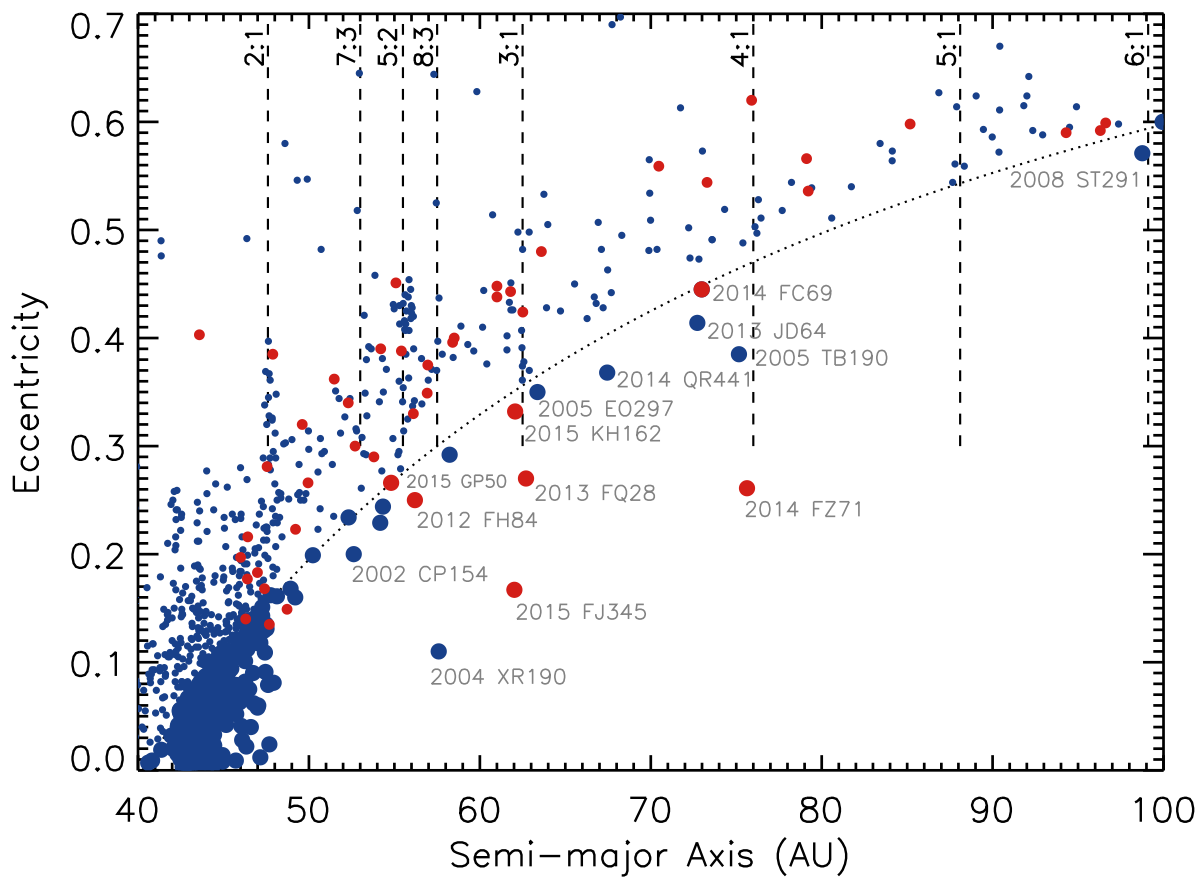}}
\caption{The semi-major axis versus eccentricity.  Red circles show
  the new objects discovered in this survey.  Larger circles show
  objects with perihelia above 40 AU.  Dashed lines show strong mean
  motion resonances with Neptune.  The dotted line shows a constant
  perihelion of 40 AU.  Objects to the right of the dotted line have
  perihelia above 40 AU and thus are mostly decoupled from Neptune.
  Uncertainties on the orbital parameters are smaller than the
  symbols.}
\label{fig:kboea2016blowup} 
\end{figure}

The new objects discovered in our survey are shown with the well known
outer solar system objects in Figure~\ref{fig:kboeq2016}.  The region
of orbital space beyond 50 AU in semi-major axis but with moderate to
low eccentricities ($e\lesssim 0.3$) has been called the Kuiper Belt
edge since only 2004 XR190 was known to occupy this area until now.
Several of our new objects have perihelia well above the generally
accepted perihelion limit where Neptune has significant influence
($q>40-41$ AU: Gomes et al. 2008; Brasser and Schwamb 2015).  Though
they have high perihelia, they only have moderate semi-major axes
($50<a<100$ AU) unlike the extreme and inner Oort cloud objects with
$a>150$ AU that likely have a different history and were detailed in
Trujillo and Sheppard (2014).  As seen in
Figure~\ref{fig:kboea2016blowup}, it appears most of these moderate
objects beyond the Kuiper Belt edge are near strong Neptune MMRs
(Table 1).  This suggests these moderate orbits were created through
MMR interactions.

This situation is similar to that of high perihelion object 2004 XR190
(Gomes 2011), though the new objects do not have exceptionally high
inclinations like 2004 XR190 ($i=46.7$). A high inclination of over 40
degrees is required for the KR mechanism to efficiently operate and
modify orbits by itself (Kozai 1962; Lidov 1962).  More moderately
inclined objects with inclinations of 20 to 40 degrees can have their
orbits significantly modified by the KR if they are also in a MMR
(Duncan \& Levison 1997; Fernandez et al. 2004).  The combined MMR+KR
mechanism could allow objects to obtain perihelia up to 60 AU (Gomes
et al. 2008).

The new objects have been observed for one to three years and thus
their orbital elements are secure.  We used the MERCURY numerical
integrator (see appendix) to look at the behaviour of all the new
objects shown in Table 1.  We found all of the new orbits to be very
stable over the age of the solar system.  As detailed later, we
examined the resonance argument angles for signs of libration, which
would indicate MMR membership (Chiang et al. 2003; Elliot et al. 2005;
Gladman et al. 2008; Pike et al. 2015).  But the objects only need to
have been in a Neptune MMR in the past to have had their orbits
significantly modified by the MMR+KR mechanism (Gallardo 2006a).  If
not in but near a MMR today, the objects could have either escaped or
Neptune migrated away to remove them from the MMR as suggested for
2004 XR190's orbit (Gomes et al. 2011).  Based on the Neptune MMR maps
shown in Gallardo (2006b), all the new very high perihelion, moderate
semi-major axis objects are near strong Neptune MMRs
(Figure~\ref{fig:kboea2016blowup}).

\subsection{The Very High Perihelion of 2014 FZ71}
One of the most interesting new objects is 2014 FZ71, which has the
highest perihelion of any known object after Sedna and 2012 VP113
(Figure~\ref{fig:kboeq2016}).  But 2014 FZ71's moderate eccentricity
and semi-major axis compared to Sedna and 2012 VP113 suggests it has a
different origin.  2014 FZ71 is very close to the 4:1 MMR with Neptune
and thus 2014 FZ71's orbit was likely modified through interactions
with it.  Interestingly, the large perihelion of 55.9 AU suggests 2014
FZ71 would not currently have any strong interaction with Neptune.
The relatively moderate inclination and eccentricity of 2014 FZ71 make
it harder to invoke the Kozai mechanism for the high perihelion of
2014 FZ71.  In our numerical simulations with ten one sigma orbit
clones, we find some of the basic 4:1 resonance argument angles,
called $e^{3}$, $es^{2}$ and $e^{2}e_{N}$ in Elliot et al. (2005),
showed signs of libration in some clones.  This indicates 2014 FZ71
likely still interacts with the 4:1 Neptune MMR.  We found all one
sigma 2014 FZ71 clones showed constant semi-major axis but some showed
large variations in $i$ and $e$ ($8<i<32$ degrees and $0.23<e<0.50$
giving $38<q<58$ AU).

If 2014 FZ71 does have both the Kozai and Neptune MMR acting on it, the
eccentricity of the object could vary and would be coupled to the
inclination following

\begin{figure}
\centerline{\includegraphics[angle=90,totalheight=0.6\textheight]{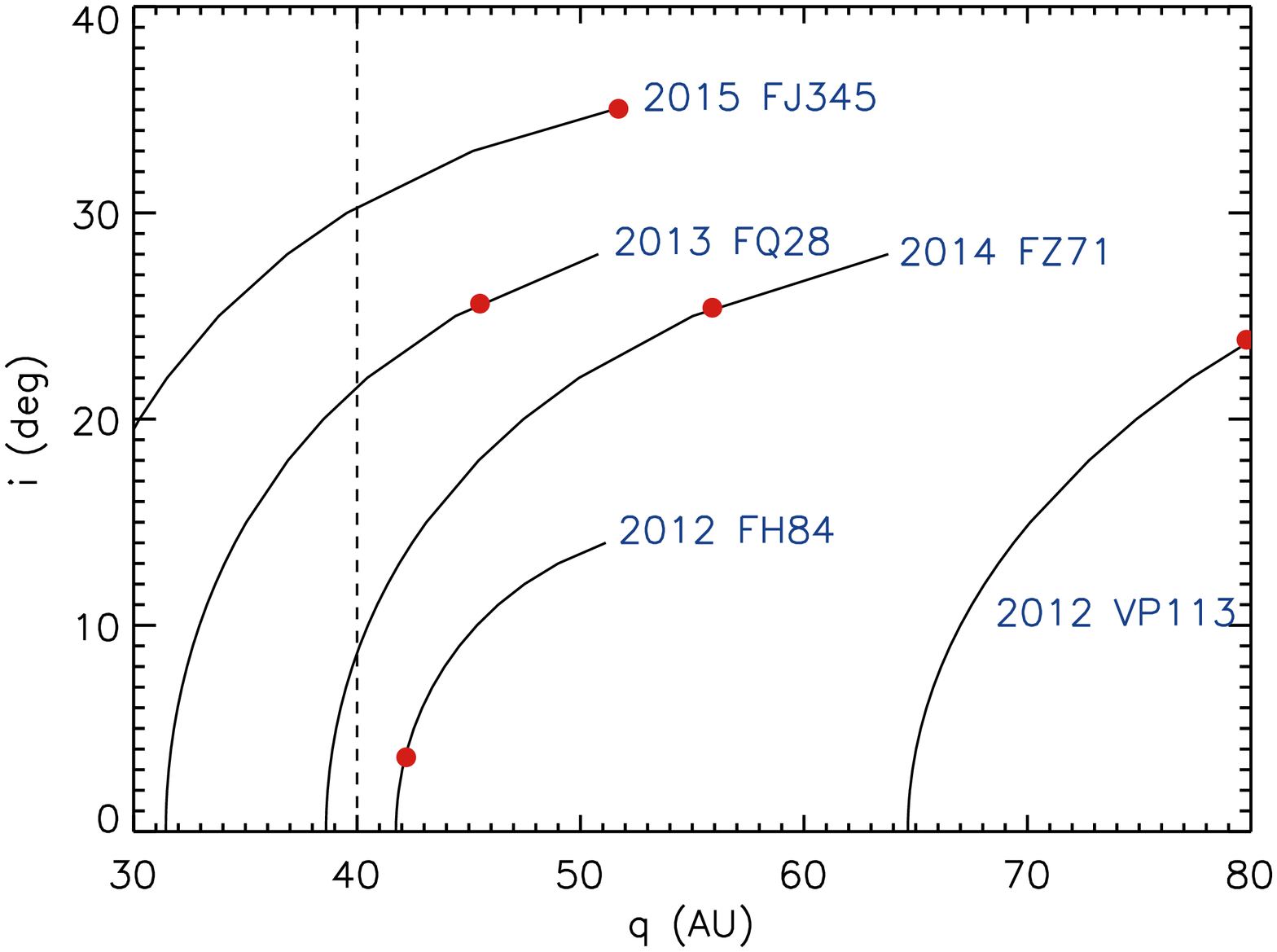}}
\caption{The MMR+KR curves for some objects.  An object that interacts
  with a Neptune MMR and the KR can have its inclination and
  perihelion (linked to eccentricity) altered through the relation $H=
  \sqrt{1-e^{2}}cos(i)$, where H is a constant and $q= a(1-e)$.  An
  object needs to come within about 40 AU for the MMR+KR to be viable.
  This fails for the extreme or inner Oort cloud objects like 2012
  VP113 or for the outer classical belt objects like 2012 FH84.  2014
  FZ71 is near the limit for the MMR+KR mechanism.  The MMR+KR
  mechanism can sufficiently explain 2015 FJ345's and 2013 FQ28's high
  perihelion orbits.}
\label{fig:kboHamiltonian} 
\end{figure}

$H= \sqrt{1-e^{2}}cos(i)$

where H is constant (Kozai 1962; Morbidelli and Thomas 1995; Gomes et
al. 2008).  In this formalism, the perihelion of 2014 FZ71 could have
been near 38.5 AU if its eccentricity was higher in the past
(Figure~\ref{fig:kboHamiltonian}).  Indeed, a perihelion of around 38
AU is exactly what we find as the lower perihelion limit for the
librating clones of 2014 FZ71 in our numerical simulations.  This
distance is just below the 40 AU upper limit Gomes et al. (2008)
suggest for the KR and Neptune MMR objects.  2014 FZ71 is an
interesting case that appears to be near the limits of effectiveness
for the MMR+KR mechanism to operate.  It is possible that 2014 FZ71 is
a more extreme case of (145480) 2005 TB190, which Gomes et al. (2008)
suggest was created by the Neptune 4:1 MMR+KR interactions.

\subsection{A Large Population of MMR+KR 3:1 Resonance Objects}
2015 FJ345, 2013 FQ28 and 2015 KH162 all have orbits near the 3:1 MMR
with Neptune.  In our numerical simulations, some of 2015 FJ345's and
2013 FQ28's one sigma clones showed oscillating resonant argument
angles with Neptune's 3:1 MMR.  2015 KH162 and its clones showed no
oscillating resonant argument angles and is thus likely a fossilized
3:1 MMR+KR object from Neptune's outward migration.  The likely 3:1
object (385607) 2005 EO297 was previously suggested by Gomes et
al. (2008) to have been created from MMR+KR interactions.  2013 FQ28
and especially 2015 FJ345 have much higher perihelia and less
eccentric orbits and thus have commonalities with 2004 XR190 and 2014
FZ71.  2015 FJ345 has the lowest eccentricity and highest perihelia of
the 3:1 objects, which is consistent with the MMR+KR being responsible
since 2015 FJ345 also has the highest inclination.  The minimum
perihelia for all these 3:1 objects could be below 35 AU through the
MMR+KR mechanism, allowing strong interactions with Neptune
(Figure~\ref{fig:kboHamiltonian}).  Our new discoveries 2015 FJ345 and
2013 FQ28 are the first 2 objects that have very high perihelia orbits
through the 3:1 Neptune MMR+KR (both also have inclinations above 25
degrees). As seen in Figure~\ref{fig:kboea2016blowup}, there is also a
cluster of objects near the 3:1 Neptune MMR with perihelia just below
40 AU.

\subsection{Other High Perihelion, Moderate Objects}

\begin{figure}
\epsscale{0.4}
\centerline{\includegraphics[angle=90,totalheight=0.6\textheight]{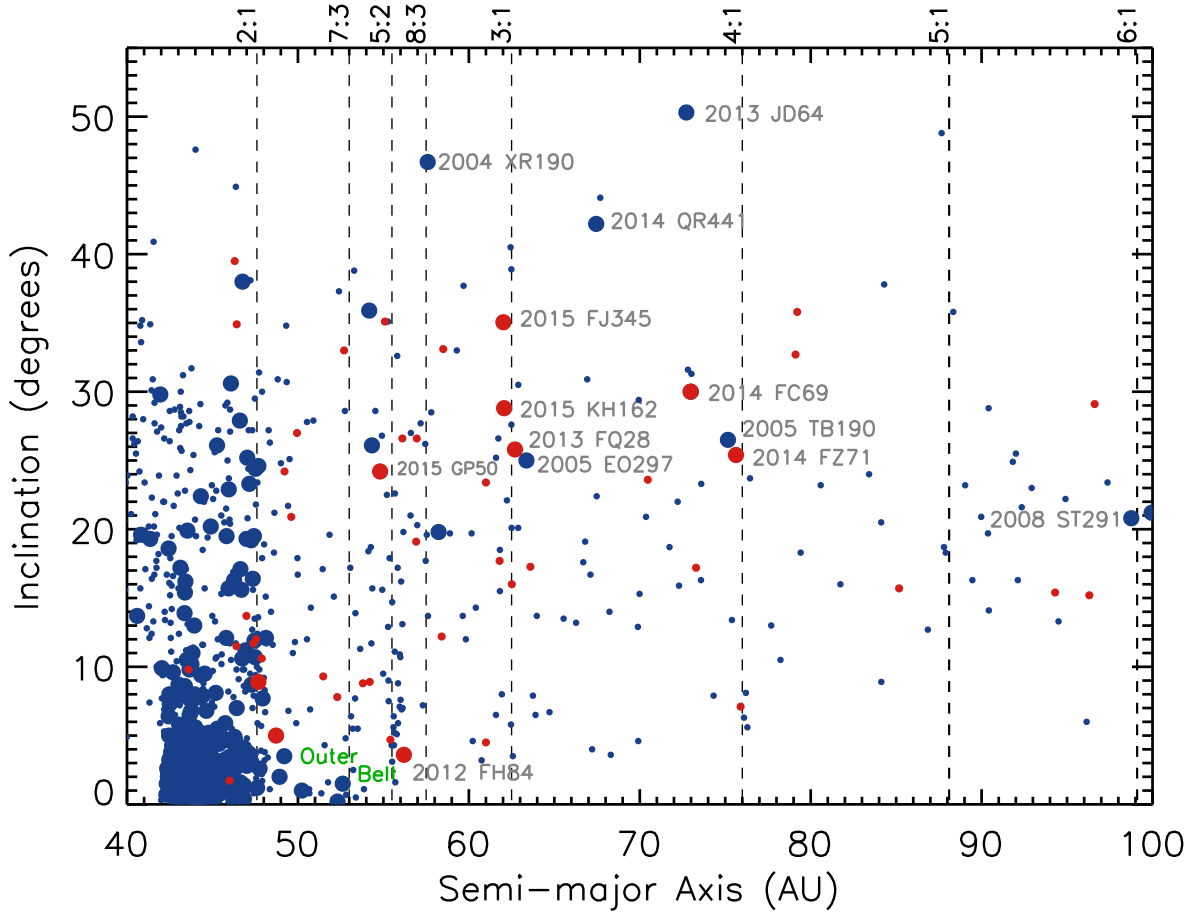}}
\caption{Similar to Figure~\ref{fig:kboea2016blowup} but showing
  inclination.  Larger circles show objects that have perihelia above
  40 AU.  All high perihelion objects beyond 53 AU have inclinations
  greater than 20 degrees except for our newly discovered 2012 FH84,
  which is likely a rare outer classical belt object along with 1995
  TL8, 2002 CP154 and 2001 FL193 (large blue circles with very low
  inclinations between 50 and 53 AU).}
\label{fig:kboia2016blowup} 
\end{figure}

There are a few objects in Figure~\ref{fig:kboia2016blowup} that have
moderately high perihelia but are not near Neptune MMRs.  The closest
resonance for 2014 FC69 and 2013 JD64 is the 11:3.  Both these objects
have very high inclinations of 30.1 and 50.3 degrees, respectively,
strongly suggesting their orbits have been created through some
interaction with the KR.  2014 QR441 is also not near any major
Neptune MMR, though the moderately strong 10:3 resonance is nearby.
2014 QR441 has a very high inclination of 42.2 degrees, again showing
the KR is likely involved.

We also find that 2008 ST291 is likely a 6:1 resonance or fossilized
resonance object that has probably been modified by the MMR+KR
mechanism.  Though the nominal orbital position does not, clones a few
tenths of AU lower in semi-major axis show resonant argument
librations (the $e^{5}$) for 2008 ST291's orbit in our numerical
simulations, where $10<i<35$ degrees, $0.40<e<0.65$ and $35<q<58$ AU
occurred over 1 Gyr.  2010 ER65 could be a similar 6:1 case, but we
found no significant resonant argument librations.

\subsection{The Outer Classical Belt}
Our new discovery 2012 FH84 also has a high perihelia and moderate
semi-major axis and eccentricity (Table 1).  But 2012 FH84 has a very
low inclination of only 3.6 degrees and is between the 5:2 and 8:3
Neptune MMRs, which makes it less likely to have been created by
MMR+KR.  Its minimum perihelion would be about 42 AU through this
mechanism, so Neptune would be unlikely to have strong interactions.
2012 FH84 is similar to 1995 TL8 ($a=52.3$ AU, $e=0.234$, $i=0.2$
deg), which cannot be explained by the MMR+KR mechanism (Gomes et
al. 2008).  2012 FH84 is thus likely a new member of the rare outer
classical belt of objects.  These are non-resonant objects that have
semi-major axes just beyond the 2:1 resonance, with moderate to low
eccentricities and low inclinations.  This outer belt might be related
to the low inclination objects in the main classical Kuiper belt as
they have similar dynamics and very red colors (Gomes et al. 2008;
Morbidelli et al. 2008; Sheppard 2010).  2002 CP154 and 2001 FL193 are
the only other objects beyond 50 AU that have perihelia higher than 40
AU and low inclinations like 2012 FH84 and 1995 TL8
(Figure~\ref{fig:kboia2016blowup}).  2014 FA72, 2013 GQ136 and 2003
UY291 are also near this region with low inclinations and perihelia
above 40 AU but have semi-major axes just below 50 AU.

The new object 2015 GP50 has a very similar semi-major axis and
eccentricity to 2012 FH84, but 2015 GP50's significantly higher
inclination could allow it to obtain a much lower perihelion.
2015 GP50 again is not obviously near a Neptune MMR but the strong 5:2
resonance is nearby. 2005 CG81's and 2007 LE38's similarly high
perihelia and highly inclined orbits, are also close to the 12:5
Neptune MMR.

\section{Discussion and Conclusions}

The moderate eccentricity space just beyond the Kuiper Belt edge at 50
AU is shown to be populated with objects other than 2004 XR190.  All
the new moderate eccentricity, very high perihelion objects ($q>45$
AU) are near strong N:1 Neptune MMRs.  We find all the moderate
eccentricity objects with perihelia above 40 AU and semi-major axes
beyond 53 AU have inclinations above 20 degrees (except the outer
classical 2012 FH84 detailed above).  Those away from Neptune N:1 MMRs
generally have the highest inclinations, which presents evidence that
the KR alone can raise the perihelion of high inclination objects
while more moderate inclinations require the addition of MMRs
(Figure~\ref{fig:kboia2016blowup}).  We used our observational bias
simulator detailed in Trujillo and Sheppard (2014) to examine the
distribution of inclinations of the MMR+KR objects in Table 1.  Using
the sin i / single Gaussian functional form for inclinations in Gulbis
et al (2010), we find the debiased inclination distribution of the
MMR+KR objects to be $\mu_1 = 28^{+2}_{-1}$ degrees and $\sigma_1 =
2.5^{+2.2}_{-0.8}$. This is significantly greater than the scattered
objects with $\mu_1 = 19.1^{+3.9}_{-3.6}$ and $\sigma_1 =
6.9^{+4.1}_{-2.7}$ (Gulbis et al. 2010).

The few colors that have been obtained for these high perihelion,
moderate orbit objects show them to be typical of scattered disk
objects (Sheppard 2010).  If these two populations of objects were
both originally from the same population, this suggests it is the
action of the MMR+KR that is responsible for the larger inclinations
seen in Table 1.  These objects were likely scattered into these
orbits and captured into resonances.  Whatever created the Kuiper Belt
edge likely occurred during or before the emplacement of MMR+KR
fossilized objects like 2004 XR190 as these fossilized objects would
likely have been lost like any other objects beyond the edge.  This
would suggest the edge was created before Neptune finished migrating
outwards and created the fossilized MMR+KR objects.

Our observational bias simulator was further used to get a crude
estimate on the MMR+KR population.  We used a uniform simulated orbit
distribution with a minimum of 0.1 eccentricity and 40 AU perihelion
with an inclination distribution described above.  We would only
detect the objects when beyond 50 AU and expect no longitudinal bias
as our survey is fairly uniform.  Because some MMRs are closer than
others we would expect population ratio detections of
1.0/0.97/0.79/0.38/0.17/0.09 for MMRs 5:2/8:3/3:1/4:1/5:1/6:1 assuming
equal populations.  The odds of finding three 3:1 high perihelion MMR
objects and no 5:2 or 8:3 objects by chance is 2.5\% if their
populations are equal (though increases to 7\% if 2015 GP50 is in the
5:2).  This suggests the 3:1 may harbor many more MMR+KR objects than
the 5:2 or 8:3 MMR, which is surprising as Volk et al. (2016) find a
large 5:2 MMR population with lower perihelia and Brasil et
al. (2014b) suggest the 3:1 and 5:2 should be the most populated with
MMR+KR objects.  However, the low order N:1 resonances like the 3:1
are the strongest for diffussing scattered objects via MMR+KR
(Gallardo 2006a).  We find that about $2400^{+1500}_{-1000}$ MMR+KR 3:1
and about $1600^{+2000}_{-1200}$ 4:1 objects larger than 100 km in
diameter likely exist with perihelia greater than 40 AU with the 5:2
and 8:3 populations significantly smaller.

Trujillo and Sheppard (2014) first noticed that the extreme
trans-Neptunian objects exhibit a clustering in their orbital angles
and predict a super-Earth planet exists beyond a few hundred AUs to
create this clustering.  Recently Batygin and Brown (2016) obtained a
possible rudimentary orbit for this planet predicted by Trujillo and
Sheppard (2014).  In our numerical integrations (see appendix) we
found this planet (a=700 AU, e=0.6 and i=30 degrees) has no
significant impact on the current MMR+KR objects, including the most
distant 2008 ST291.  We note that all five our new MMR+KR objects
along with 2004 XR190 have longitudes of perihelion ($LP = \omega +
\Omega$) between about 80 and 190 degrees, which is about 180 degrees
from the longitude of perihelion for the ETNOs.

\section*{Acknowledgments}

We thank Y. Ramanjooloo and D. Hung for help in recovery of 2015 KH162
at the Hawaii 88inch telescope.  This project used data obtained with
the Dark Energy Camera (DECam), which was constructed by the Dark
Energy Survey (DES) collaboration. Funding for the DES Projects has
been provided by the DOE and NSF (USA), MISE (Spain), STFC (UK), HEFCE
(UK). NCSA (UIUC), KICP (U. Chicago), CCAPP (Ohio State), MIFPA (Texas
A\&M), CNPQ, FAPERJ, FINEP (Brazil), MINECO (Spain), DFG (Germany) and
the collaborating institutions in the DES, which are Argonne Lab, UC
Santa Cruz, University of Cambridge, CIEMAT-Madrid, University of
Chicago, University College London, DES-Brazil Consortium, University
of Edinburgh, ETH Zurich, Fermilab, University of Illinois, ICE
(IEEC-CSIC), IFAE Barcelona, Lawrence Berkeley Lab, LMU Munchen and
the associated Excellence Cluster Uni verse, University of Michigan,
NOAO, University of Nottingham, Ohio State University, University of
Pennsylvania, University of Portsmouth, SLAC National Lab, Stanford
University, University of Sussex, and Texas A\&M University.  Based in
part on observations at CTIO, NOAO, which is operated by AURA under a
cooperative agreement with the NSF.  Based in part on data collected
at Subaru Telescope, which is operated by the National Astronomical
Observatory of Japan.  C.T. was supported by the Gemini Observatory.
This research was funded by NASA grant NN15AF446. This paper includes
data gathered with the 6.5 meter Magellan Telescopes located at Las
Campanas Observatory, Chile.

\appendix
\section{Appendix}

Our simple numerical simulations were performed in order to determine
the basic orbital properties and behaviour of the newly discovered
objects.  We used the MERCURY numerical integrator (Chambers 1999).
In our basic simulations we used the four giant planets Jupiter,
Saturn, Uranus and Neptune and added the mass of the terrestrial
planets to the Sun.  An additional simulation was run with all the
same conditions but adding in a 15 Earth mass planet on an eccentric
$e=0.6$ orbit at 700 AU to see how it might effect the orbits of the
MMR+KR objects.  The time step used was 20 days and all integrations
ran for over 1 billion years.  Orbital elements used were heliocentric
converted from the barycentric output from the orbit fitting program
by Bernstein and Khushalani (2000).  In our simulations of the nominal
orbits and ten clones within 1 sigma of each new object's orbit we
found no significant semi-major axis variability over 1 billion years.
For most nominal orbits and clones the $e$ for all the new objects
only varied by 0.01 to 0.02 and the $i$ at most by about 3 degs over 1
billion years.  But some 1 sigma clones did show large variations in
$e$ and $i$ indicating significant interactions with Neptune's MMRs.
2014 FZ71 with a tenth of an AU larger semi-major axis than the
nominal position showed variations of $8<i<32$ degrees, $0.50>e>0.23$
inversely with $i$ and $38<q<58$ AU over 100 Myr timescales,
indicating interactions with Neptune's 4:1 MMR.  A tenth of an AU
smaller clone of 2013 FQ28 near the 3:1 Neptune MMR had $i$ vary from
20 to 30 degrees and $e$ inversely from 0.2 to 0.4 giving a perihelion
from 38 to 50 AU over 1 billion years.  The 2008 ST291's clones of a
few tenths of an AU smaller than the nominal position showed
significant orbital variability in $e$ and $i$.  2008 ST291's clones
in the Neptune 6:1 MMR resonance where the MMR+KR mechanism allowed
$i$ to vary from 35 to 10 degrees and $e$ inversely from 0.40 to 0.65
over 100 Myrs (with perihelia ranging between 35 to 58 AU).  Including
the distant massive planet didn't cause the clones of 2008 ST291 to
escape the 6:1 Neptune MMR or 2014 FZ71 to escape the 4:1 Neptune MMR
and their basic orbital behaviour was similar to the simulations
without the putative distant massive planet.

\newpage

\begin{center}
\begin{deluxetable}{lccccccccccc}
\tablenum{1}
\tablewidth{6.8 in}
\tablecaption{New High Perihelion Objects with Moderate Semi-Major Axes and Eccentricities}
\tablecolumns{12}
\tablehead{
\colhead{Name} & \colhead{$q$}  &  \colhead{$a$} & \colhead{$e$}  & \colhead{$i$} & \colhead{$\Omega$} & \colhead{$\omega$} &  \colhead{Dist}   & \colhead{Dia}  & \colhead{$m_{r}$} & \colhead{N} & \colhead{$R:R$} \\ \colhead{} & \colhead{(AU)} & \colhead{(AU)}  & \colhead{} &\colhead{(deg)} &\colhead{(deg)} & \colhead{(deg)} & \colhead{(AU)}  & \colhead{(km)} & \colhead{(mag)} & \colhead{} & \colhead{} }  
\startdata
\multicolumn{12}{c}{\textbf{Neptune MMR + KR}}  \nl
2014 FZ71      &   55.9   &   76.4   &    0.268  &    25.440 &    305.85   &    243.7  & 56.8    &   150    &   24.4  & 12  &   4:1       \nl
2015 FJ345     &   51.8   &   62.5   &    0.17   &    35.00  &    37.88    &    80.4   & 58.5    &   100    &   25.5  & 13  &   3:1      \nl
\textit{2004 XR190} & \textit{51.2} & \textit{57.5} & \textit{0.110} & \textit{46.7} & \textit{252.4} & \textit{283.4} & \textit{58.5} & \textit{600} & \textit{21.8} & \textit{-} & \textit{8:3} \nl
2013 FQ28      &   45.8   &   63.2   &    0.27   &    25.70  &    214.89   &    230.4  & 66.8    &   250    &   24.1  & 15  &   3:1       \nl
\textit{2008 ST291*} & \textit{42.3} & \textit{98.8} & \textit{0.572} & \textit{20.8} & \textit{324.2} & \textit{331.2} & \textit{56.7} & \textit{600} & \textit{21.5} & \textit{-} & \textit{6:1} \nl
2015 KH162     &   41.5   &   62.1   &    0.33   &    28.8   &    200.8    &    296.1  & 58.8    &   800    &   21.1  & 41  &   3:1      \nl
2015 GP50      &   40.5   &   55.3   &    0.27   &    24.15  &    222.69   &    128.4  & 68.2    &   200    &   24.7  & 13  &   5:2?     \nl
2014 FC69      &   40.5   &   72.9   &    0.44   &    30.1   &    250.2    &    189.3  & 83.7    &   500    &   23.6  & 10  &   11:3?      \nl 
\multicolumn{12}{c}{\textbf{Outer Classical Belt}} \nl
2012 FH84       &   42.7   &   56.4   &    0.24   &    3.62   &    21.37    &    7.2    & 68.1    &   150    &   25.3  &  9  &   5:2?    \nl
\enddata
\tablenotetext{}{Objects in italics were known before this work and * shows a new result. Quantities are the perihelion ($q$), semi-major axis ($a$), eccentricity ($e$), inclination ($i$), longitude of the ascending node ($\Omega$), argument of perihelion ($\omega$), distance (Dist), number of observations (N), and Neptune resonance ($R:R$). Diameter (Dia) assumes a moderate albedo of 0.10. Orbits are from the MPC and uncertainties are shown by the number of significant digits.}
\end{deluxetable}
\end{center}

\newpage

\newpage

\newpage

\newpage

\begin{references}

\reference{All02} Allen, R., Bernstein, G., \& Malhotra, R. 2002, AJ, 124, 2949.

\reference{All06} Allen, R., Gladman, G., Kavelaars, J., Petit, J., Parker, J. \& Nicholson, P. 2006, ApJ, 640, L83.

\reference{Bat16} Batygin, K. \& Brown, M. 2016, AJ, 151, 22.

\reference{Ber00} Bernstein, G. \& Khushalani, B. 2000, AJ, 120, 3323.

\reference{Bra14a} Brasil, P., Nesvorny, D., \& Gomes, R. 2014a, AJ, 148, 56.

\reference{Bra14b} Brasil, P., Gomes, R., \& Soares, J. 2014b, AA, 564, A44.

\reference{Bra12} Brasser, R., Duncan, M., Levison, H., Schwamb, M. \& Brown, M. 2012, Icarus, 217, 1.

\reference{Bra15} Brasser, R. \& Schwamb, M. 2015, MNRAS, 446, 3788.

\reference{Bro04} Brown, M., Trujillo, C. \& Rabinowitz, D. 2004, ApJ, 617, 645.

\reference{Cha99} Chambers, J. 1999, MNRAS, 304, 793.

\reference{Chi03} Chiang, E., Jordan, A., Millis, R. et al. 2003, AJ, 126, 430.

\reference{Daw12} Dawson, R. \& Murray-Clay, R. 2012, ApJ, 750, 43.

\reference{Dun97} Duncan, M. \& Levison, H. 1997, Science, 276, 1670.

\reference{Ell05} Elliot, J., Kern, S., Clancy, K., et al. 2005, AJ, 129, 1117.

\reference{Fla15} Flaugher, B., Diehl, H., Honscheid, K., et al. 2015, AJ, 150, 150.

\reference{Fer04} Fernandez, J., Gallardo, T. \& Brunini, A. 2004, Icarus, 172, 372.

\reference{Gal06b} Gallardo, T. 2006a, Icarus, 181, 205.

\reference{Gal06a} Gallardo, T. 2006b, Icarus, 184, 29.

\reference{Gal12} Gallardo, T., Hugo, G. \& Pais, P. 2012, Icarus, 220, 392.

\reference{Gla02} Gladman, B., Holman, M., Grav, T., Kavelaars, J., Nicholson, P., Aksnes, K. \& Petit, J. 2002, Icarus, 157, 269.

\reference{Gla06} Gladman, B. \& Chan, C. 2006, ApJ, 643, L135.

\reference{Gla08} Gladman, B., Marsden, B. \& VanLaerhoven, C. 2008, in The Solar System Beyond Neptune, eds. M. Barucci, H. Boehnhardt, D. Cruikshank and A. Morbidelli (Tucson: Univ of Arizona Press), 43-57.

\reference{Gla12} Gladman, B., Lawler, S., Petit, J. et al. 2012, AJ, 144, 23.

\reference{Gom05} Gomes, R., Gallardo, T., Fernandez, J., \& Bruini, A. 2005, Celest. Mech. Dynam. Astron., 91, 109.

\reference{Gom06} Gomes, R., Matese, J., \& Lissauer, J. 2006, Icarus, 184, 589.

\reference{Gom08} Gomes, R., Fernandez, J., Gallardo, T. and Brunini, A. 2008, in The Solar System Beyond Neptune, eds. M. Barucci, H. Boehnhardt, D. Cruikshank and A. Morbidelli (Tucson: Univ of Arizona Press), 259-273.

\reference{Gom11} Gomes, R. 2011, Icarus, 215, 661.

\reference{Gul10} Gulbis, A., Elliot, J., Adams, E., Benecchi, S., Buie, M., Trilling, D. \& Wasserman, L. 2010, AJ, 140, 350.

\reference{Hol93} Holman, M. \& Wisdom, J. 1993, AJ, 105, 1987.

\reference{Jew98} Jewitt, D., Luu, J., \& Trujillo, C. 1998, AJ, 115, 2125.

\reference{Key04} Kenyon, S. \& Bromley, B. 2004, Nature, 432, 598.

\reference{Koz62} Kozai, Y. 1962, AJ, 67, 591.

\reference{Lid62} Lidov, M. 1962, P\&SS, 9, 719.

\reference{Mal95} Malhotra, R. 1995, AJ, 110, 420.

\reference{Mor95} Morbidelli, A. and Thomas, F. 1995, Icarus, 118, 322.

\reference{Mor04} Morbidelli, A. \& Levison, H. 2004, AJ, 128, 2564.

\reference{Mor08} Morbidelli, A., Levison, H. and Gomes, R.  2008, in The Solar System Beyond Neptune, eds. M. Barucci, H. Boehnhardt, D. Cruikshank and A. Morbidelli (Tucson: Univ of Arizona Press), 275-292.

\reference{Nes15a} Nesvorny, D. 2015a, AJ, 150, 68.

\reference{Nes15b} Nesvorny, D. 2015b, AJ, 150, 73.

\reference{Nes16} Nesvorny, D. \& Vokrouhlicky, D. 2016, arXiv160206988.

\reference{Pet11} Petit, J., Kavelaars, J., Gladman, B. et al. 2011, AJ, 142, 131.

\reference{Pik15} Pike, R., Kavelaars, J., Petit, J., Gladman, B., Alexandersen, M., Volk, K. \& Shankman, C. 2015, AJ, 149, 202.

\reference{Sch10} Schwamb, M., Brown, M., Rabinowitz, D. \& Ragozzine, D. 2010, ApJ, 720, 1691.

\reference{She10} Sheppard, S. 2010, AJ, 139, 1394.

\reference{She12} Sheppard, S. 2012, AJ, 144, 169.

\reference{Soa13} Soares, J., and Gomes, R. 2013, AA, 553, 110.

\reference{Tor90} Torbett, M. \& Smoluchowski, A. 1990, Nature, 345, 49.

\reference{Tru01} Trujillo, C. \& Brown, M. 2001, ApJ, 554, L95.

\reference{Tru14} Trujillo, C. \& Sheppard, S. 2014, Nature, 507, 471.

\reference{Vol16} Volk, K., Murray-Clay, R., Gladman, B. et al. 2016, arXiv:1604:08177.

\end{references}
\end{document}